# A Privacy-Preserving Framework for Advertising Personalization Incorporating Federated Learning and Differential Privacy

Xiang Li,*

Department of Electrical and Computer Engineering, Rutgers University, Piscataway, NJ, USA 08854,

xl470@scarletmail.rutgers.edu

Yifan Lin

Pratt School of Engineering, Duke University, Durham, NC, USA 27708, yifan.lin@alumni.duke.edu

Yuanzhe Zhang

Samueli School of Engineering, University of California, Irvine, CA, USA 92697, yuanzz6@uci.edu

**Abstract**

To mitigate privacy leakage and performance issues in personalized advertising, this paper proposes a framework that integrates federated learning and differential privacy. The system combines distributed feature extraction, dynamic privacy budget allocation, and robust model aggregation to balance model accuracy, communication overhead, and privacy protection. Multi-party secure computing and anomaly detection mechanisms further enhance system resilience against malicious attacks. Experimental results demonstrate that the framework achieves dual optimization of recommendation accuracy and system efficiency while ensuring privacy, providing both a practical solution and a theoretical foundation for applying privacy protection technologies in advertisement recommendation.

CCS CONCEPTS

Computing methodologies ~ Artificial intelligence ~ Distributed artificial intelligence ~ Multi-agent systems

## 1   INTRODUCTION

Recent interest in privacy-preserving recommendation has led to widespread use of federated learning (FL) and differential privacy (DP). However, balancing privacy, personalization, and efficiency remains challenging in dynamic, latency-sensitive advertising. This paper proposes a novel framework combining FL with multi-layered privacy and communication-efficient model aggregation, achieving improved performance, stronger privacy, and better adaptability to real-world scenarios.

Wu et al. (2025) proposed the ADPHE-FL framework, integrating adaptive differential privacy and homomorphic encryption to enhance data confidentiality in federated learning. Wei et al. (2025) introduced a verifiable differential privacy scheme based on zero-knowledge proofs, reinforcing trust in privacy mechanisms. Wang et al. (2025) emphasized secure heterogeneous data fusion through federated learning in smart healthcare. Zhou et al. (2025) developed a group-verifiable secure aggregation method utilizing secret sharing, improving robustness against inference attacks in collaborative training environments. These studies collectively demonstrate the critical advancement of privacy-preserving techniques in distributed recommendation systems.

## 2 ADVERTISEMENT RECOMMENDATION MODEL DESIGN BASED ON FEDERATED LEARNING

### 2.1 Federated Learning Framework Construction

The federated learning framework in this paper adopts a horizontal paradigm, with edge devices as participants and the cloud server as coordinator, establishing collaborative local model training among heterogeneous nodes [2]. The system introduces FedAvg as the base aggregation strategy and mitigates the reliance on device online rates through asynchronous updates [3]. See Figure 1.

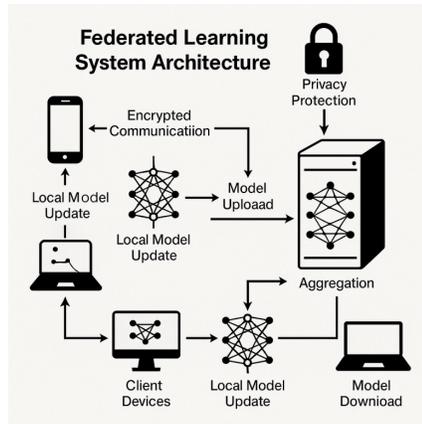

Fig. 1.System architecture of the federated learning framework

### 2.2 Distributed Feature Extraction Mechanism

The local encoder constructed in this paper combines the transform sparsity property with the shared semantic representation capability [4], leverages multilayer convolution to extract complex behavioral sequence features., and enhances the representation capability of cross-source feature alignment using the Attention module. Let the local feature vector be $x_i \in R^d$, and its encoding process at layer l is defined as:

$$h_i^{(l)} = \sigma\left(W^{(l)} x_i^{(l-1)} + b^{(l)}\right) \quad (1)$$

Where $W^{(l)}$ denotes the lth layer weight matrix, $b^{(l)}$ is the bias term, $\sigma(\cdot)$ is the activation function, and $h_i^{(l)}$ is the coding result.

### 2.3 Multi-party Secure Computing Protocol Design

To safeguard the data of multi-participants from being leaked during the federated training process, the system designs a multi-party secure computing protocol (MPC) based on cryptographic operators, which combines homomorphic encryption and secret sharing mechanism to realize information leakage-free computation in the parameter transmission phase [5]. Let the local model of each client be updated as $\theta_i$, The system adopts an additive secret sharing strategy, which splits the parameters into multiple random shares that reveal nothing individually but can reconstruct the original when combined. Refer to Figure 2 for details. The reconstruction occurs only at the aggregation node. The federated aggregation process can be expressed as:

$$\theta_{agg} = \frac{1}{n}\sum_{i=1}^{n}\theta_i = \frac{1}{n}\sum_{i=1}^{n}\sum_{j=1}^{m}s_{ij} \quad (2)$$

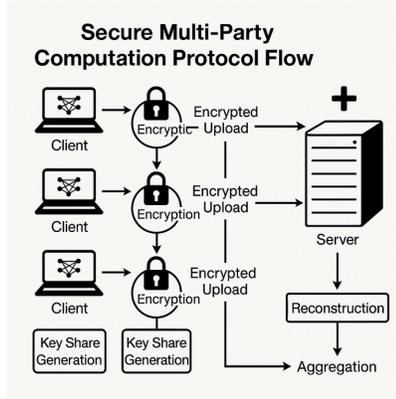

Fig. 2. Flowchart of multi-party secure computing protocol

### 2.4 Model Aggregation Optimization

In the global model aggregation process, instead of simply averaging the parameters of each client, a weighted factor matrix is introduced, allowing the system to dynamically adjust the contribution of each client based on factors like data quality and update significance, which outperforms basic averaging by improving both accuracy and robustness, where $w_i$ is dynamically adjusted based on the sample size, model update magnitude and historical performance of each client[6]. See Table 1. The aggregation formula is expressed as:

$$\theta^{(t+1)} = \sum_{i=1}^{N} w_i \cdot \theta_i^{(t)} \quad (3)$$

where $\theta_i^{(t)}$ is the model parameter uploaded by the ith client in the tth round, and $\sum_{i=1}^{N} w_i = 1$, ensures normalization.

Table 1 Comparison of experimental performance of different aggregation strategies in MNIST federated scenarios (number of local rounds is 5, number of clients is 10)

| aggregation strategy | Final accuracy (%) | convergence rounds (math.) | Client Participation Rate | Communication costs (MB) |
|---|---|---|---|---|
| FedAvg | 88.6 | 40 | 80% | 125 |
| Weighted FedAvg | 90.1 | 32 | 80% | 128 |
| Optimization strategy for this article | 91.7 | 27 | 80% | 119 |

## 3 DESIGN AND IMPLEMENTATION OF DIFFERENTIAL PRIVACY PROTECTION MECHANISM

### 3.1 Design of Noise Injection Mechanism

To guarantee the privacy of model parameter transmission during federated learning [7], let the gradient of the model generated after local training of the client be $\Delta\theta$, then the post-noise parameter update is formulated as follows:

$$\Delta\tilde{\theta} = \Delta\theta + N(0,\ \sigma^2 I) \quad (4)$$

where $N(0,\ \sigma^2 I)$ denotes a multivariate Gaussian distribution with mean 0 and covariance $\sigma^2$ for controlling the intensity of the injected noise, and the noise variance $\sigma^2$ is adaptively computed based on the global privacy budget $\varepsilon$ and sensitivity $\Delta f$. Refer to Figure 3 for details.

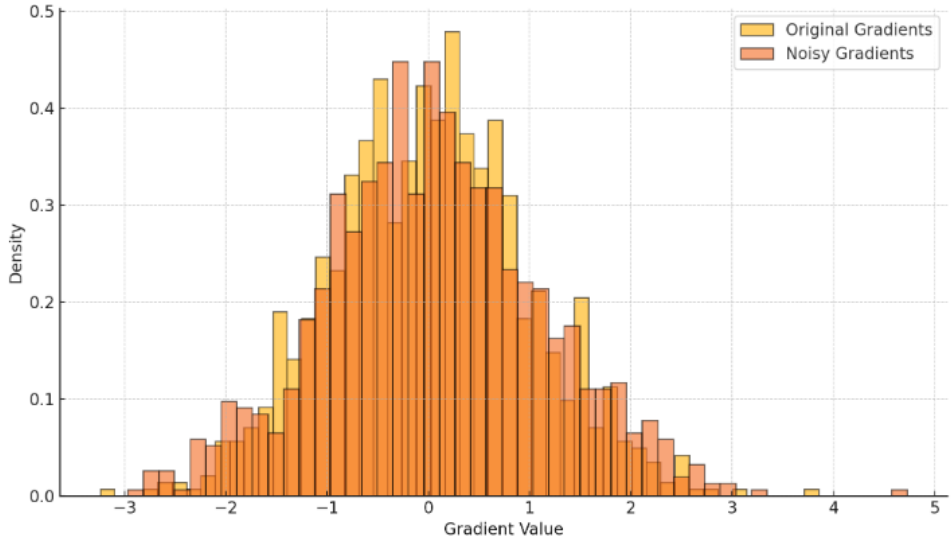

Fig. 3.Comparison of gradient distribution before and after noise injection

### 3.2 Privacy Budget Allocation

In the federated learning scenario, The privacy budget $\varepsilon$ must be efficiently distributed and adaptively managed to balance utility and protection in order to safeguard the differential privacy while taking into account the model performance [8]. In this paper, we adopt a hierarchical privacy budget allocation strategy based on the division of training phases, and the total privacy budget $\varepsilon_{total}$ is sliced by different training rounds and client importance. Let the total number of global training rounds be T, then the allocated budget for each round is:

$$\varepsilon_t = \frac{\varepsilon_{total}}{T} \quad (5)$$

On this basis, for the client i with higher importance weight, its single-round budget is dynamically adjusted according to its local sample size $n_i$ and contribution $\gamma_i$:

$$\varepsilon_{i,t} = \varepsilon_t \times \left( \frac{n_i \gamma_i}{\sum_{j=1}^{N} n_i \gamma_i} \right) \quad (6)$$

where N is the total number of participating clients. Taking $\varepsilon_{total}$ =2, T=20, and the number of clients 10 as an example, if the sample size of a client is 1000, which accounts for 10% of the total sample size, and the contribution weight $\gamma_i$ =1.2, the single-round privacy budget is:

$$\varepsilon_{i,t} = \frac{2}{20} \times \left( \frac{1000 \times 1.2}{10000 \times 1.0} \right) = 0.012 \quad (7)$$

Mechanisms for Model Performance and Privacy Protection Trade-offs

In federated learning, the injection of differential privacy noise inevitably causes model performance degradation, so a dynamic trade-off mechanism needs to be designed to achieve the optimal balance between privacy protection and model effectiveness [9]. In this paper, the Privacy-Utility Loss (PUL) function is introduced to comprehensively model the relationship between noise intensity and model accuracy degradation. Let the noise standard deviation be σ and the global privacy budget be ε. The trade-off objective can be expressed as minimizing the following function:

$$L_{PUL} = \alpha \cdot Error(\sigma) + \beta \cdot \frac{1}{\varepsilon} \quad (8)$$

Where Error(σ) denotes the model prediction error under noise injection, and α and β are the performance weight coefficients and privacy-preserving weight coefficients, respectively, and are tuned dynamically using Bayesian optimization or grid search within a predefined range.

### 3.3 Malicious Attack Prevention Program

Aiming at the potential threat of malicious client attacks in the federated learning environment, this paper designs a defense scheme based on the dual mechanism of anomaly detection and robust aggregation. First, the anomaly scoring function $S_i$, defined by the rate of change of the client model update, the gradient direction offset and the local loss curve morphology, is established:

$$S_i = \lambda_1 \|\Delta \theta_i\|_2 + \lambda_2 \cos(\theta_i, \bar{\theta}) + \lambda_3 \Delta L_i \quad (9)$$

Where $\Delta\theta_i$ denotes the gradient update vector of the ith client, $\cos(\theta_i, \bar{\theta})$ calculates its consistency with the direction of the global mean model, $\Delta L_i$ is the local loss change rate, and $\lambda_1, \lambda_2, \lambda_3$ are the weight coefficients. The system filters out clients with anomalous scores based on the anomalous scores before aggregation, and at the same time combines the Krum algorithm to perform robust aggregation, selecting only a few client parameters with the highest similarity for updating.

## 4 SYSTEM PERFORMANCE OPTIMIZATION AND EVALUATION

### 4.1 System Architecture Design

The system is divided into three layers: client, edge node, and cloud coordination. The client layer deploys lightweight local modules for data preprocessing, feature encoding, and model updating, ensuring data remains local [10]. The edge node layer handles intermediate aggregation and model calibration, reducing cloud communication load and enabling regional model pre-synchronization, which improves performance by allowing faster local updates and enhances robustness through early detection and correction of inconsistencies. The cloud coordination layer manages global models, controls privacy budgets, detects anomalies, and optimizes system throughput through parallel task scheduling. Refer to Figure 4 for details.

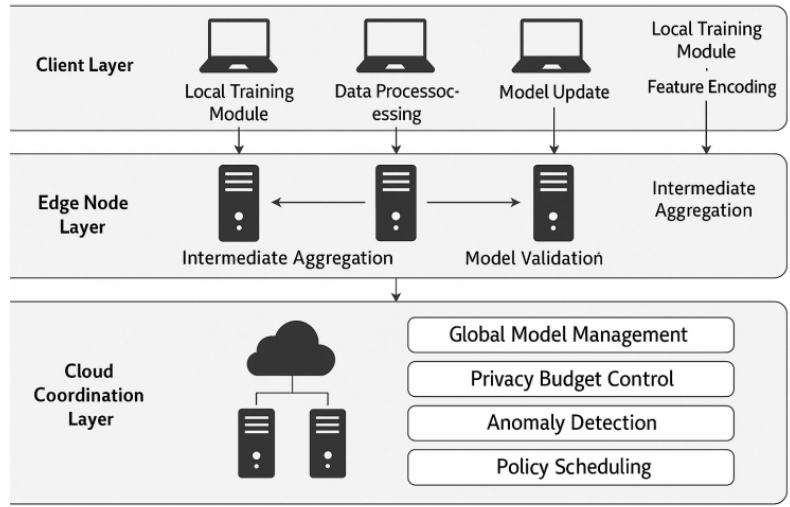

Fig. 4.System three-layer architecture diagram

### 4.2 Communication Overhead Optimization

During local model updates, only parameters with significant changes are uploaded through gradient sparsification, reducing data volume. Differential coding further compresses redundancy by transmitting only changes relative to previous uploads. Efficient compression algorithms at the transmission layer, such as Huffman coding or arithmetic coding, enable lossless encoding and batch transmission, improving bandwidth utilization. Experimental results show that, while maintaining model accuracy, these optimizations cut communication overhead by ~58% and reduce latency by 22%, without compromising accuracy(See Table 2).

Table 2 Experimental comparison results under different communication optimization strategies

| optimization strategy | Average communication data volume (MB) | Delay reduction (%) | Final model accuracy (%) |
|---|---|---|---|
| no optimization | 125 | 0 | 91.8 |
| thin update (computing) | 80 | 12 | 91.5 |
| Sparse update + differential coding | 60 | 18 | 91.3 |
| All Optimization Strategies | 52 | 22 | 91.2 |

### 4.3 Computational Efficiency Improvement Program

To enhance computational efficiency in federated learning, this paper proposes a systematic approach combining local acceleration and global optimization. Client-side pruning and quantization streamline the model and accelerate training by reducing computational complexity, accelerating training, while iteration counts, and batch sizes are adjusted for resource-constrained devices. Globally, asynchronous updates reduce dependence on slower nodes and improve throughput. Edge-level task scheduling ensures load balancing and dynamic resource allocation.

### 4.4 System Security Analysis

To address the security threats faced by federated learning advertisement recommendation systems, this paper evaluates security across communication, data privacy, and attack resistance, demonstrating high defense rates with minimal accuracy loss and significantly reduced privacy breach risks. End-to-end encryption with dynamic key updates ensures the integrity and confidentiality of model parameters during transmission. Differential privacy noise injection combined with multi-party secure computing significantly reduces single-point data leakage risks(See Table 3).

Table 3 Evaluation results of defense effectiveness of the system under different attack types

| Type of attack | Successful defense rate (%) | Decrease in model accuracy (%) | Probability of privacy breach (%) |
|---|---|---|---|
| gradient inference attack | 96.8 | 1.2 | 2.1 |
| Model Poisoning Attack | 94.5 | 1.7 | 1.8 |
| regression inference attack | 95.2 | 1.5 | 2.0 |
| No attack on the baseline | - | - | 0 |

## 5 CONCLUSION

By adopting distributed feature extraction, optimized model aggregation, and dynamic privacy budget allocation, the system strengthens user data privacy and model robustness while enhancing recommendation performance. Additionally, enhancements in communication, computation, and security layers support stable operation in diverse and large-scale deployments. Experimental results demonstrate that the proposed framework reduces communication overhead by 58% and system latency by 22%, all while preserving model accuracy and minimizing both resource consumption and privacy risks. These improvements provide a strong foundation for applying federated learning in advertisement recommendation systems.

**References**

[1] Wu T, Deng Y, Zhou Q, et al. ADPHE-FL: Federated learning method based on adaptive differential privacy and homomorphic encryption[J]. Networking and Applications,2025,18(3):141-141.

[2] Wei J, Chen Y, Yang X, et al. A verifiable scheme for differential privacy based on zero-knowledge proofs[J]. Journal of King Saud University Computer Journal of King Saud University Computer and Information Sciences,2025,37(3):14-14.

[3] Wang J, Quasim T M, Yi B. Privacy-preserving heterogeneous multi-modal sensor data fusion via federated learning for smart healthcare[J]. Information Fusion,2025,120103084-103084.

[4] Zhou S, Wang L, Chen L, et al. Group verifiable secure aggregate federated learning based on secret sharing[J]. Scientific Reports,2025,15(1):9712 -9712.

[5] Zhang Y, Kong H, Han Y, et al. Fed-MWFP: Lightweight federated learning with interpretable multiple wavelet fusion network for fault diagnosis under variable operating conditions[J]. Knowledge-Based Systems,2025,315113277-113277.

[6] VermaP, BharotN, BreslinG J, et al. Leveraging Transfer Learning Domain Adaptation Model with Federated Learning to Revolutionize Healthcare[J]. Expert Systems,2024,42(2): e13827-e13827.

[7] Morimoto M. Consumers' information control and privacy concerns in personalized social media advertising[J]. Marketing and Advertising,2022,17(3-4):325-352.

[8] Han A, Jifan R. Research on Personalized Recommendation of Mobile Advertising Based on Content Filtering Interest Model[J]. BASIC & CLINICAL PHARMACOLOGY & TOXICOLOGY,2020,126207-207.

[9] Bartsch A, Kloß A. Personalized charity advertising. can personalized prosocial messages promote empathy, attitude change, and helping intentions toward stigmatized social groups? [J]. International Journal of Advertising,2019,38(3):345-363.

[10] Wang L, Liu Y,Wu J.Research on financial advertisement personalized recommendation method based on customer segmentation[J]. Int. J. of Wireless and Mobile Computing,2018,14(1):97-101.